\documentstyle[prb,aps,multicol,epsf]{revtex}
 \tighten \begin{document}
\draft
\title{Charged surface in salty water with multivalent ions:
Giant inversion of charge.}
 \author
{T. T. Nguyen, A. Yu. Grosberg, and B. I. Shklovskii}
\address{Department of Physics, University of Minnesota, 116 Church
St. Southeast, Minneapolis, Minnesota 55455} \maketitle
\begin{abstract}
Screening of a strongly charged macroion by
oppositely charged colloidal particles, micelles,
or short polyelectrolytes is considered.
Due to strong lateral repulsion such multivalent counterions
form a strongly correlated liquid at the surface of the macroion.
This liquid provides correlation induced attraction of
multivalent counterions to the macroion surface.
As a result even a moderate concentration of multivalent
counterions in the solution inverts the sign of the net macroion charge.
We show that at high concentration of monovalent salt
the absolute value of inverted charge can be
larger than the bare
one. This giant inversion of charge can be observed in electrophoresis.

\end{abstract}


 \begin{multicols}{2}

Charge inversion is a phenomenon in which
a charged particle (a
macroion) strongly binds so many counterions
in a water solution, that its net charge
changes sign. As shown below the binding energy of counterion
with large charge $Z$ is larger than 
$k_B T$, so that this
 net charge is easily observable; for instance, it is
the net charge that determines linear transport properties, such
as particle drift in a weak field electrophoresis.  Charge
inversion has been observed~\cite{Dubin} in polyelectrolyte-micelle
system and is possible for a variety of other systems, ranging
from solid surface of mica or lipid membranes, to DNA or actin.

Charge inversion is of special interest
for delivery of genes to the living cell for the purpose of gene
therapy. The problem is that both bare DNA and a cell
surface are negatively charged and repel each other. The goal is
to screen DNA in such a way that the resulting complex
is positive~\cite{Felgner}.

Theoretically, charge inversion can be also thought of as an
over-screening.  Indeed, the simplest screening atmosphere,
familiar from linear Debye-H\"{u}ckel theory, compensates at any
finite distance only a part of the macroion charge.  It can be
proven that this property holds also in non-linear
Poisson-Boltzmann (PB) theory.
The statement that the net charge
preserves sign of the bare charge agrees with the common sense.
One can think that this statement is even more universal than
results of PB equation.  It was
shown~\cite{Roland,Perel99,Shklov99}, however, that this
presumption of common sense fails for screening by $Z$-valent
counterions ($Z$-ions), such as charged colloidal particles, micelles,
or short polyelectrolytes, because there are strong lateral
correlations between them when they
are bound to the surface of a macroion.  These correlations are
beyond the mean field PB theory, and charge inversion is their
most spectacular manifestation.

Charge inversion has attracted a significant attention in the last
couple of years~\cite{Pincus}.  Our goal in the present paper is to
provide a simple physical explanation of charge inversion and to
show that in the most practical case, when both
$Z$-ions and monovalent salt, such as NaCl, are present,
 not only charge
sign may flip, but the inverted charge can become
even larger in absolute value than the bare charge, 
thus giving rise to
{\it giant charge inversion}.

Let us demonstrate the role of lateral
correlations between $Z$-ions for a
primitive toy model. Imagine
a hard-core sphere with radius $b$ and
with negative charge $Q$ screened
by two spherical positive $Z$-ions with radius $a$.
One can see that if Coulomb repulsion between $Z$-ions
is much larger than $k_BT$ they are situated
on opposite sides of the negative sphere (Fig. 1a).
If $Ze < 2|Q|$ each $Z$-ion is bound, because
the energy required to remove it
to infinity $|Q|Ze/(a+b) - Z^{2}e^{2}/2(a+b)$ is positive.
Thus, the charge of the whole complex $Q + 2Ze$
can be positive and
as large as $3|Q|$. This example demonstrates
the possibility of an
almost 300\% charge inversion. It is obvious that
this charge inversion is a result of the correlation between
$Z$-ions which avoid each other and
reside on opposite sides of the negative charge.
On the other hand, description of screening of the central sphere
in PB approximation smears the positive charge,
as shown on Fig. 1b and does not lead to the charge inversion. Indeed,
in this case charge accumulates in spherically symmetric screening
atmosphere only until the point of neutrality at which electric field
reverses its sign and attraction is replaced by repulsion.

\begin{figure}
\epsfxsize=5cm \centerline{\epsfbox[150 500 500 720]{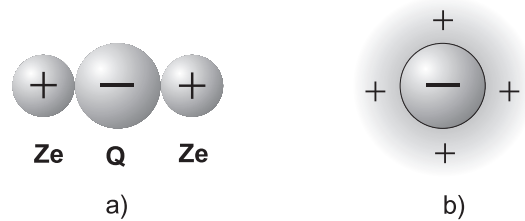}}
\caption{a) A toy model of charge inversion.
b) PB approximation does not lead to charge inversion.}
\end{figure}

In this paper we consider screening of a macroion
surface with negative immobile
surface charge density $-\sigma$
by finite concentration of positive $Z$-ions, neutralizing amount 
of monovalent coions, and a large
concentration $N_1$ of a monovalent salt.
This is more practical problem than one considered in  
Ref.~\onlinecite{Perel99,Shklov99}, where
monovalent salt was absent.
Correspondingly, we assume that all interactions are screened with
Debye-H\"{u}ckel screening length
$r_s = \left(8\pi l_{B}N_1\right)^{-1/2}$,
where $l_{B} = e^{2}/(Dk_B T)$ is the Bjerrum length,
$e$ is the charge of a proton, $D \simeq 80$
is the dielectric constant of water.

We begin with the simplest macroion which is a thin charged sheet
immersed in water solution (Fig. 2a). 
Later we examine more realistic macroion which is a thick insulator 
charged at the surface (Fig. 2b).

\begin{figure}
\epsfxsize=6cm \centerline{\epsfbox[0 0 450 380]{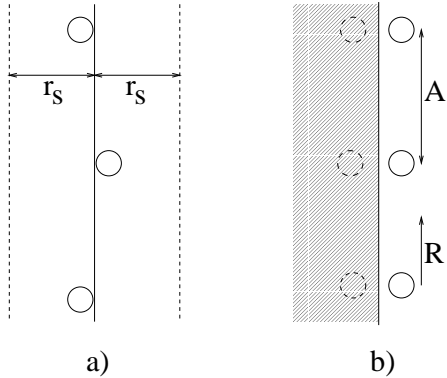}}
\vspace{0.2cm}
\caption{Models studied in this paper. $Z$-ions are shown
by full circles. a) Charged plane
immersed in water.  b)Surface of a large macroion. Image
charges are shown by broken circles.}
\end{figure}

Assume that the plane with the charge density $-\sigma$
is covered by $Z$-ions with two-dimensional concentration $n$.
Integrating out all monovalent ions, or,
equivalently, considering all interactions screened at the distance
$r_s$, we can write down the free energy per unit area in the form
\begin{equation}
F = \pi \sigma^{2} r_s/D  -2 \pi \sigma r_s Zen/D +  F_{ZZ} + F_{id},
\label{free}
\end{equation}
where the four terms are responsible, respectively, for
the self interaction of the charged plane, 
for the interaction  between
$Z$-ions and the plane, for the interaction
between $Z$-ions and for the entropy of ideal two-dimensional gas 
$Z$-ions.

Our goal is to calculate the net charge density of the plane
\begin{equation}
\sigma^* = -\sigma + Zen.
\label{sigma*}
\end{equation}
Using Eq.~(\ref{sigma*}) one can rewrite Eq.~(\ref{free}) as
\begin{equation}
F= \pi (\sigma^{*})^{2} r_s/D + F_{OCP},
\label{free1}
\end{equation}
where $F_{OCP} = F_{c}+ F_{id}$ is the free energy
 of the same system of $Z$-ions
residing on a neutralizing background with surface charge density
$-Zen$, which is conventionally referred to as
one component plazma (OCP), and
\begin{equation}
F_{c}= -\pi (Zen)^2 r_s/D  + F_{ZZ}.
\label{free2}
\end{equation}
is the correlation part of $F_{OCP}$. 
This transformation can be simply interpreted as the
addition of uniform charge densities $-\sigma^*$
and $\sigma^*$ to the plane.
The first addition makes a neutral OCP on the plane.
The second plane of charge creates two plane capacitors
with negative charges on both sides of the plane which screen
inverted charge of the plane at the distance $r_s$.
The first term of Eq.~(\ref{free1})
is nothing but the energy of these two capacitors.
There is no cross term in energy between the OCP and the capacitors
because each plane capacitor creates a constant potential,
$\psi(0) = 2\pi \sigma^{*} r_s/D$, at the neutral OCP.

Using Eq.~(\ref{free2}), the electrochemical potential of $Z$-ions
 at the plane can be written as
$\mu = Ze \psi(0) + \mu_{id}  + \mu_{c}$,
where $\mu_{id}$ and $\mu_{c}=\partial F_{c}/\partial n$ are
the ideal and the correlation parts of the chemical potential of OCP.
In equilibrium, $\mu$ is equal to the chemical
potential, $\mu_{b}$ of the bulk
solution,
because in the bulk electrostatic potential $\psi=0$.
Using Eq. (\ref{free1}), we have:
\begin{equation}
2\pi \sigma^{*} r_s Ze/D = - \mu_{c} + (\mu_{b}- \mu_{id})
\label{master}
\end{equation}
As we show below, in most practical cases the correlation effect
is rather strong, so that $\mu_{c}$ is negative and $|\mu_{c}|\gg k_BT$.
This means that for large enough concentration of $Z$-ions in the
bulk and at the surface, $n$, both bulk chemical 
potential $\mu_{b}$ and
ideal part of surface chemical potential $\mu_{id}$ should be
neglected compared to $\mu_{c}$. 
Furthermore, strong correlations
imply that at least short range order of $Z$-ions on the surface
should be similar to that of triangular Wigner crystal (WC)
since it delivers the lowest energy to OCP. Therefore,
\begin{equation}
\sigma^{*} = \frac{D}{2 \pi r_s} \frac{\left| \mu_{c} \right|}{Z
e} \simeq \frac{D}{2 \pi r_s} \frac{\left| \mu_{WC} \right|}{Z e}\ . 
\label{capacitor}
\end{equation}
We see now that the net charge density 
$\sigma^{*}$ is positive. This proves
inversion of the bare charge density $-\sigma$.
Eq. (\ref{capacitor}) has a very simple meaning: 
$|\mu_{WC}|/Ze$
is the "correlation" voltage which charges two above mentioned
parallel capacitors with thickness $r_s$ and total
capacitance per unit area $D/(2\pi r_s)$.

To calculate the "correlation" voltage $\left| \mu_{WC} \right|
/Ze$, we start from the case of weak screening when $r_s$ 
is larger than the average distance between $Z$-ions.
In this case, screening does not affect thermodynamic properties of
WC. The energy per $Z$-ion $\varepsilon(n)$
of such Coulomb WC at $T=0$ can be estimated as an 
interaction energy of a $Z$-ion
with its Wigner-Seitz cell, because interaction 
energy of neigboring neutral Wigner-Seitz cells
is very small.
This gives $\varepsilon(n)=- Z ^{2}e^{2}/R
D$, where $R=(\pi n)^{-1/2}$ is
the radius of a Wigner-Seitz cell (we approximate hexagon by a disc).
More accurately~\cite{mara}
$\varepsilon(n) = -1.1Z ^{2}e^{2}/R D = - 1.96 n^{1/2}Z^{2}e^{2}/D.$
One can discuss the role of a finite temperature on WC  in terms of 
the inverse dimensionless temperature
$\Gamma = Z^{2}e^{2}/(R D k_BT)$. We are 
interested in the case of large $\Gamma$.
For example, at a typical $Zen = \sigma = 1.0~e/$nm$^{2}$ and at
room temperature, $\Gamma = 10$ even for $Z=4$.
Wigner crystal melts~\cite{Gann} at  $\Gamma = 130$,
so that for $\Gamma < 130$ we deal with a
strongly correlated liquid. Numerical
calculations, however, confirm that
at $\Gamma \gg 1$ thermodynamic properties of strongly correlated
liquid are close to that of WC~\cite{Totsuji}.
Therefore, for estimates of $\mu_{c}$
we can still write that $F_{c} = n\varepsilon(n)$ 
and use
\begin{equation}
\mu_{WC} = \frac{\partial\left(n\varepsilon(n)\right)}{\partial n}
= - 1.65\Gamma k_BT = -1.65 \frac {Z^{2}e^{2}}{R D}.
\label{muwc}
\end{equation}
We see now that indeed $\mu_{WC}$ is negative and $|\mu_{WC}| \gg k_BT$,
so that Eq.~(\ref{capacitor}) is justified.
Substituting Eq.~(\ref{muwc}) into Eq.~(\ref{capacitor}),
we get $\sigma^{*} = 0.83 Ze/(\pi r_s R)$.
At $r_s \gg R$, charge density $\sigma^{*}\ll \sigma$,
 and $Zen \simeq \sigma$,
one can replace $R$ by $R_0= (\sigma\pi/Ze)^{-1/2}$.
This gives
\begin{equation}
\sigma^{*}/\sigma =  0.83 (R_0/r_s) = 0.83 \zeta^{1/2},~~~(\zeta \ll 1)
\label{smally}
\end{equation}
where $\zeta = Ze/\pi \sigma r_s^2$ is a
dimensionless charge of a $Z$-ion.
Thus, at $r_s \gg R$ or $\zeta \ll 1$, inverted charge
density grows with decreasing
$r_s$. Extrapolating to $r_s = 2R_0$ where
screening starts to substantially modify
the interaction between $Z$-ions
we obtain $\sigma^{*}=0.4\sigma$.

Now we switch to the case of strong screening, $r_s \ll R$, or
$\zeta \gg 1$. It seems that in this case $\sigma^{*}$
should decrease with decreasing  $r_s$, because screening
reduces the energy of WC and leads to its melting. In fact, this
is what eventually happens. However, there is
a range of $r_s \ll R$ where the energy of WC is still large.
In this range, as $r_s$ decreases, the repulsion between
$Z$-ions becomes weaker,
 what in turn makes it easier to
pack more of them on the plane.
Therefore, $\sigma^{*}$ continues to grow with
decreasing $r_s$.

At $r_s \ll R$ one is still able to estimate
thermodynamic properties of OCP from the model
of a triangular WC.
Keeping only interactions with the 6 nearest neighbors
in Eq. (\ref{free2}), we can write the correlation part of
free energy of screened WC per unit area as
\begin{equation}
F_{c}= - \frac{\pi r_s(Zen)^{2}}{D}  +
3n\frac{(Ze)^{2}}{DA}\exp(-A/r_s),
\label{SCWC}
\end{equation}
where $A = (2/\sqrt3)^{1/2}n^{-1/2}$
is the lattice constant of this WC.
Calculating the chemical potential of $Z$-ions
at the plane, $\mu_{WC} = \partial F_{c}/\partial n$ and substituting it
into Eq.~(\ref{capacitor})
one finds that $A\simeq r_s\ln(3\zeta/4)$,
$R \simeq (2\pi/\sqrt3)^{1/2}r_s\ln(3\zeta/4)$ and
\begin{equation}
\frac{\sigma^{*}}{\sigma}
=\frac{2\pi\zeta}{\sqrt3~ \ln^{2}(3\zeta/4)} -1,~~~(\zeta\gg 1).
\label{giant}
\end{equation}
Alternatively, one can derive  Eq.~(\ref{giant})
by direct minimization of
Eq.~(\ref{free}) with respect of $n$. In this way, one 
does not need a capacitor interpretation which is not 
as transparent in this case as for $r_s \gg R$.
 
Thus, at $r_s \ll R$, or
$\zeta \gg 1$ the distance $R$ decreases and 
inverted charge continues
to grow with decreasing $r_s$. This result
could be anticipated for the toy model of Fig. 1a if 
Coulomb interaction betwen the spheres is
replaced by a strongly screened one.
Screening obviously affects repulsion between positive spheres
stronger than their attraction to the negative one
and, therefore, makes maximum allowed charges $Ze$ larger.

Above we studied analytically
two extremes,
$r_s \gg R$ and $r_s \ll R$. In the case of arbitrary $r_s$
we can find $\sigma^{*}$ numerically.
For this purpose we calculate $\mu_{WC}$
from Eq.~(\ref{free2}) and substitute it in
Eq.~(\ref{capacitor}). This gives
\begin{equation}
\frac{1}{\zeta}=\sum_{\bbox{r}_i \neq 0}
\frac{3+r_i/r_s}{8~r_i/r_s} e^{-r_i/r_s}~~~,
\label{exacty}
\end{equation}
where the sum is taken over all vectors of WC lattice
and can be evaluated numerically. Then
one can find the equilibrium $n$ for
any given values of $\zeta$. The resulting ratio $\sigma^*/\sigma$
is plotted by the solid curve in Fig. 3.

\begin{figure}
\epsfxsize=4 cm \centerline{\epsfbox[180 510 300 670]{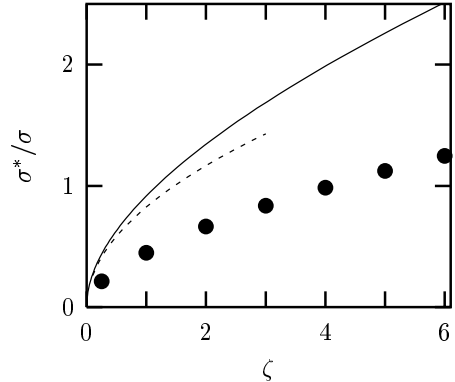}}
\caption{The ratio $\sigma^*/\sigma$ as a function
of the charge $\zeta$.
The solid curve is calculated for a charged plane by
a numerical solution to eq. (\ref{exacty}),
the dashed curve
is the large $r_s$ limit, eq. (\ref{smally}). The $\bullet$ points
are calculated for the screening of the surface of the semispace
 with dielectric constant
much smaller than 80.  In this case
image charges (Fig. 2b) are taken into account.
} \end{figure}

Since the value of $\sigma^{\ast}$ represents the main result of
our work, its subtle physical meaning should be clearly
understood.  Indeed, the entire system, macroion plus overcharging
$Z$-ions, is of course neutralized by the monovalent salt.
One can ask then, what is the meaning of charge inversion?  The
answer is simple for $r_s \gg R$, when charge $\sigma^{*}$ is well
separated in space from the oppositely charged atmosphere of
monovalent salt (which leads to the 
interpretation  based on two capacitors, see
above).  When $r_s \ll R$ there is no such obvious spatial
separation. Nevertheless, $\sigma^{*}$ can be observed, because
$Z$-ions are bound with energies well above $k_B T$ while
small ions are only weakly bound. First, the number of bound
$Z$-ions can be counted using, e.g., the atomic force
microscopy. Positive $\sigma^{*}$  means "over-population": there
are more bound $Z$-ions than neutrality condition implies.
Second, it is $\sigma^{*}$ that determines the mobility of
macroion in the weak field electrophoresis experiments.

The results discussed so far were derived for the charged plane
which is immersed in water and screened on both sides by
$Z$-ions and monovalent salt (Fig. 2a).  In reality
charged plane is typically a surface of a rather thick membrane
whose (organic, fatty) material is a dielectric with permeability
much less than that of water. In this case, image
charges which have the same sign as $Z$-ions must be
taken into account (Fig. 2b). We have analyzed
this situation in details,
which will be reported elsewhere.  The main result turns out to be
very simple: while image charges repel $Z$-ions and drive the
entire Wigner crystal somewhat away from the surface, their major
effect is that in this case only one capacitor must be charged (on
the water side of the surface). Accordingly, the ratio
$\sigma^{*}/\sigma$ is reduced by a factor very close to 2 compared
to the case of two-sided plane (Fig. 3).

We are prepared to address now the question of maximal possible
charge inversion. How far can a macroion be overcharged, and what
should one do to achieve that? Figure 3 and equation (9) suggest
that the ratio $\sigma^*/\sigma$ continues to grow with growing
$\zeta$. However, the possibilities to increase $\zeta$ are
limited along with the assumptions of the presented theory.
Indeed, there are two ways to increase $\zeta = Z e / \sigma \pi
r_s^2$, namely to choose surface with small $\sigma$ and ions with
large $Z$. The former way is restricted because $Z$-ions remain
strongly bound to the surface only as long as $\left| \mu_{WC}
\right| \simeq 2\pi r_s \sigma Z e /D \gg k_BT$ or $\zeta < 2 Z^2
l_B / r_s$.  Therefore, the latter way, which is to increase $Z$,
is really the most important.  It is, however, also restricted,
because at large $Z$, monovalent ions start to condense on the
$Z$-ion~\cite{Gueron}.  Assuming $Z$-ions are spheres of the
radius $a$, their effective net charge at large $Z$ can be written
as $Z_{\rm eff} = \left(a / l _B \right) 2 \ln \left( Z l_B r_s /
a^2 \right)$, yielding $\zeta < 8 \left( a^2 / l_B r_s \right)
\left[ \ln \left( Z l_B r_s / a^2 \right) \right]^2$. Since this
estimate was derived under the assumption that  $r_s > a$, the
largest $a$ we can choose is $a =r_s$. For $r_s = a = 10 \AA$
charge $\zeta$ may be as high as about 10, so that the ratio
$\sigma^*/\sigma$ can exceed 100\%.

Since charge inversion grows with increasing $a$ we are tempted to 
explore the case $a > r_s$.  To address this situation, our theory
needs a couple of modifications.  Specifically, in the
first term of Eq.~(\ref{SCWC}) we must take into account the fact
that only a part of $Z$-ion interacts with the surface, namely the
segment which is within the distance $r_s$ from the surface.  One
should also take into account that strong screening increases
$Z_{\rm eff}$.  Assuming $Z$-ion is a sphere, this modifies upper
bound for $\zeta$ by a factor $a/r_s$ and thus it makes charge
inversion even larger. We do not discuss this regime in details,
because it is highly non-universal, dependent on the shape and
charge distribution of the $Z$-ions, plane roughness, etc.

Meanwhile, there is much more powerful way to increase charge
inversion. Suppose we take $Z$-ions with the shape of long rigid
rods. Such a situation is very practical, since it corresponds to
the screening of charged surface by rigid polyelectrolytes, such
as DNA double helix \cite{Yang}.  In this case, correlation
between $Z$-ions leads to parallel, nematic-like ordering
of rods on the surface.  In other words, WC in this case is
one-dimensional, perpendicular to rods.  Chemical potential
$\left| \mu _{WC} \right|$ in this case is about the interaction 
energy of one rod with the stripe of the surface charge,
which plays the role of the Wigner-Seitz cell.  Importantly, this
energy, along with the effective net charge, $Z_{\rm eff}$, are
proportional to the rod length $L$ and thus can be very large.
Rods can be strongly bound, with chemical potential much exceeding
$k_B T$, even at very small $\sigma$.   This holds even in
spite of the Onsager-Manning condensation~\cite{Manning} of
monovalent ions on the rods: for instance, at $A > r_s > a$ one
has $Z_{eff} = L \eta_c/e$ , where $A$ and $a$ are, respectively,
the distance between rods in WC and radius of the rod (double
helix), $\eta_c= k_BT/e$.  As a result the
 ratio $\sigma^*/\sigma$ grows with
decreasing $r_s$ as $\sigma^*/\sigma \simeq (\eta_c/2 r_s  \sigma)
\ln \left(\eta_c/2\pi r_s  \sigma \right)$. At $r_s \sim a$ and
small enough $\sigma$ this ratio can be much larger than one. This
phenomenon can be called {\it giant charge inversion}.

Giant charge inversion can be also achieved if 
DNA screens a positively charged wide cylinder with the radius
greater or about the DNA double helix persistence length ($500
\AA$).  In this case DNA spirals around the cylinder, once again
with WC type strong correlations between subsequent turns.   We
leave open the possibility to speculate on the relevance of this
model system to the fact that DNA  overcharges a nucleosome by
about 20\%~\cite{Pincus}.

To conclude, we have presented simple physical arguments
explaining the nature and limitations of charge inversion in the
system, where no interactions are operational except for Coulomb
and short range hard core repulsion. Correlations between bound ions,
which are strong for multivalent counterions with $Z \gg 1$, are
the powerful source of charge inversion for purely
electrostatic system.  We have shown that even spherical $Z$-ions
adsorbed on a large plane macroion can lead to charge inversion 
larger than 100\%, while for rod-like $Z$-ions charge inversion can 
reach gigantic proportions.

%--------------------------------------------------------------------%
%\acknowledgements

We are grateful to R. Podgornik for attracting our interest 
to the problem of DNA adsorption on a charged surface 
and I. Rouzina for useful discussions. 
This work was supported by NSF DMR-9985985.

\end{multicols}
\end{document}